\documentclass[%
 reprint,
 amsmath,amssymb,
 aps,
]{revtex4-2}

\usepackage{graphicx}
\usepackage{dcolumn}
\usepackage{bm}
\usepackage{textcomp}

\begin{document}

\preprint{APS/123-QED}

\title{Ultra-broadband SPDC for spectrally far separated photon pairs}

\author{Aron Vanselow$^{1,3}$, Paul Kaufmann$^1$, Helen M. Chrzanowski$^1$, and Sven Ramelow$^{1,2,4}$}
 \affiliation{%
  $^1$Institute of Physics,\\
  Humboldt-Universit\"at zu Berlin, Berlin, 12489, Germany
  \\$^2$IRIS Adlershof,\\
  Humboldt-Universit\"at zu Berlin, Berlin, 12489, Germany
  \\$^3$aron.vanselow@physik.hu-berlin.de
  \\$^4$sven.ramelow@physik.hu-berlin.de
 }%




\date{\today}

\begin{abstract}
Generating photon pairs via spontaneous parametric down-conversion (SPDC) in nonlinear crystals is important for a wide range of quantum optics experiments with spectral properties such as their bandwidths often being a crucial concern. Here, we show the generic existence of particular phase-matching conditions in quasi-phase matched KTP, MgO:LN and SLT crystals that lead to ultra-broadband, widely non-degenerate photon pairs. It is based on the existence of group-velocity matched, far apart wavelength pairs and for 2 mm long crystals results in SPDC bandwidths between 15 and 25\,THz (FWHM) for photon pairs with the idler photon in the technologically relevant mid-IR range 3-5\,\textmu m and the signal photon in the NIR below 1100\,nm. We experimentally demonstrate this type of broadband phase-matching in ppKTP crystals for photon pairs centered at 800\,nm and 3800\,nm and measure a bandwidth of 15\,THz. This novel method of generating broadband photon-pairs will be highly beneficial for SPDC-based imaging, spectroscopy, refractometry and OCT with undetected mid-IR photons.
\end{abstract}

\setlength{\parskip}{0pt}
\maketitle


Over the last few decades, spontaneous parametric down-conversion (SPDC) has served as the standard workhorse for the generation of photon pairs in numerous fundamental experiments and demonstrations of photonic quantum technologies, which now touch upon relevant real-world applications. A standard technique to implement SPDC is using a second order nonlinear material, typically a nonlinear crystal, pumped by a laser to generate pairs of photons. The photons' central wavelengths and bandwidths depend on the spectrum of the pump laser and the phase-matching characteristics of the nonlinear crystal. Beside standard phase-matching techniques like birefringent phase-matching, quasi-phase-matched (QPM) MgO-doped Lithium Niobate (MgO:LN), potassium titanyl phosphate (KTP) and stoichiometric lithium tantalate (SLT) crystals are widely used due to the flexibility offered by QPM and the possibility to access the largest component of these crystals' nonlinearity tensor. 

For a number of quantum optics applications relying on SPDC, a large bandwidth of the generated photon pairs is crucial. The spectro-temporal modes of broadband SPDC have recently become a topic of renewed interest \cite{Brecht:2015}, with the spectro-temporal entanglement of bi-photons naturally yielding a large alphabet for quantum communication and quantum key distribution \cite{Roslund14}. Broadband SPDC sources are in addition highly relevant for frequency multiplexing \cite{puigibert17,Joshi:2018} which enables multiplexing into a single spectral mode without current limitations of spatial mode switching.

Other examples where widely separated, broadband photon pairs would be highly desirable are recently developed SPDC-based methods with undetected photons for imaging \cite{Lemos:2014gw}, spectroscopy \cite{Kalashnikov:2016cl}, refractometry \cite{Paterova:2015} and optical coherence tomography (OCT) \cite{Valles:2018df,Paterova:2018kl, vanselow192}. Specifically, these techniques allow one to measure optical properties of an object at the idler wavelength, while only detecting photons at the signal wavelength. Thus no detectors or laser systems at the idler wavelength are necessary. This entails a significant technological advantage when the idler wavelength is chosen in a regime where both (detectors and lasers) are technologically problematic. An example of this are measurements in the mid-IR regime, which offer the highest potential for useful applications of these schemes by choosing the signal wavelength in a spectral region like the VIS or NIR where high-performance detectors, cameras, spectrometers and optical components have orders of magnitude better performance, at a fraction of the cost.

For spectroscopy and refractometry, a broad bandwidth is desirable because it enables the investigation of the properties of the sample over a wider spectral range. For OCT, a broad bandwidth is essential because the axial resolution is inversely proportional to the spectral width. Furthermore, a broad bandwidth enables spectrally resolved imaging (hyper-spectral imaging) and thus label-free functional imaging \cite{baker14}.

In general, in order to generate very broadband SPDC, different techniques can be used. The most straightforward method is to use very short crystals, which, however, comes at the expense of a severely reduced total brightness which scales inversely with the crystal length. Another method is to use degenerate quasi-phase matching (type I or type 0) which automatically leads to broad spectra, which can be further broadened via the use of chirped quasi-phase matching \cite{Nasr:2008gu}. However, a chirped grating again comes at the price of strongly reduced brightness.

To increase the spectral bandwidth in general for second-order nonlinear interactions like second-harmonic generation (SHG) or difference- and sum-frequency generation (DFG, SFG) several other approaches have also been reported previously. One approach is to alternate the quasi-phase matching period along the length of the crystal. In \cite{Bortz:1994}, this allowed for an increase in bandwidth by a factor of 15 in Lithium Niobate (LN) waveguides, achieving bandwidths of approximately 4\,nm in the 920\,nm regime for the degenerate case. In \cite{Fujioka:2005}, non-collinear interaction is combined with degenerate quasi-phase matching and group-velocity matching for SHG. In \cite{Yanagawa:2005fb}, broadband near-degenerate type-0 difference frequency generation in LN is proposed using quasi-phase matching and group-velocity matching. They expect an idler bandwidth of 500\,nm in the 2\,\textmu m region. In several other publications, e.g. in \cite{Prakash:2008bk}, collinear near-degenerate broadband interactions using quasi-phase matching and group velocity matching have been proposed. In \cite{Brida:2007} an approach similar to the one discussed in the present paper was used for broadband optical parametric amplification of signal pulses, matching signal and idler group velocities in LiIO$_3$. They investigate numerically the resulting spectra at a single fixed pump wavelength of 800\,nm for several crystals. In \cite{Jeon:2016}, similar measurements are performed for LN using parametric amplification in the communication band. In \cite{Zhong:2015fn}, non-degenerate broadband type-0 and type-I phase matching is proposed, matching the group velocities of pump and signal. 

Here, we show ultra-broadband phase matching for SPDC that relies on group-velocity matching widely non-degenerate signal and idler wavelengths, realized by collinear type-0 quasi-phase matching. It is widely tunable through tuning of the pump and the crystal temperature. We give numerical examples for the three most widely used crystals KTP, MgO:LN and SLT, with quasi-phasematching enabling flexibility in the choice of idler (or signal) wavelength. These demonstrate the possibility to generically generate large bandwidths exceeding 15\,THz for a wide range of central mid-IR wavelengths between 3 and 5\,\textmu m for suitably chosen signal and pump wavelengths and corresponding poling periods. As mentioned above, this is a very important spectral regime. Furthermore, we experimentally confirm our predictions with experimental results for a 2\,mm long ppKTP crystal phase-matched for collinear SPDC with photon pairs centered at 800\,nm and 3800\,nm, achieving an ultra-large bandwidth of 15\,THz. The method is patented \cite{ramelow18}.

To understand the possibility for ultra-large bandwidths at such widely non-degenerate signal and idler wavelengths, it is instructive to re-derive how the SPDC bandwidth depends on the crystal dispersion properties: 
For quasi-phasematched SPDC, the pump (p), signal (s) and idler (i) frequencies obey energy conservation: $\omega_s + \omega_i = \omega_p$. The normalized spectrum of the signal/idler photons $I(\omega_{s/i})$ for a crystal of length $L$ and a monochromatic pump (p) as a function of the phase-mismatch $\Delta k$ is given by \cite{Fedrizzi13}:
\begin{equation}\label{sinc}
I(\omega_{s/i})=\text{sinc}^2(\Delta k(\omega_{s/i})\cdot L/2)
\end{equation}
with $\Delta k$ defined by
\begin{equation}\label{DeltaK}
\Delta k =k_p - k_s - k_i - \dfrac{2\pi m}{\Lambda},
\end{equation}
where $\Lambda$ is the poling period. $m$ is the QPM order which is an odd integer number. Here, $m=1$ for maximum efficiency.
\begin{figure} [htbp]
\centering
\includegraphics[width=\linewidth]{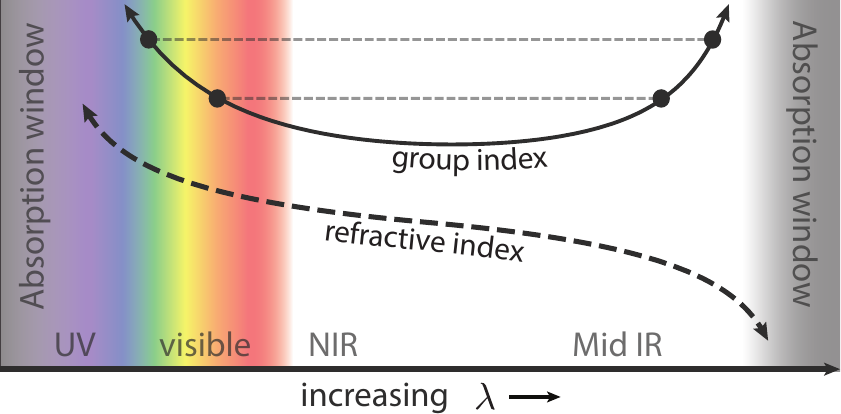}
\caption{The typical behavior of refractive index $n$ and group index $n_g$ of common nonlinear crystals like KTP, MgO:LN and SLT that yields the potential to match the group velocities of widely non-degenerate signal and idler photons. }\label{fig:MgOLNnGroup}
\end{figure}

Using the crystal dispersion $n(\omega)$, which can be modelled by crystal-specific Sellmeier equations given in the form of $n(\lambda)$, one can express $\Delta k$ as:
\begin{equation}\label{QPM}
\Delta k=\dfrac{1}{c_0} (n_p(\omega_p)\omega_p - n_s(\omega_s)\omega_s -n_i(\omega_i)\omega_i -\dfrac{1}{\Lambda}).
\end{equation}
 
For given pump, signal and idler wavelengths one can explicitly calculate $\Lambda$ to achieve phase matching where $\Delta k=0$: 

\begin{align}\label{Poling Length}
\Lambda= c_0   \cdot (n_p(\omega_p)\omega_p - n_s(\omega_s)\omega_s -n_i(\omega_i)\omega_i)^{-1}.
\end{align}

In order to maximize the bandwidth of the idler and thus also the signal photons, it is necessary to minimize the change of $\Delta k$ for a fixed pump frequency when changing the idler frequency. In first order this is equivalent to requiring
\begin{align}\label{vgMatching}
\partial \Delta k/\partial\omega_i|_{\omega_{i_0}}\stackrel{!}{=}0
\end{align}
where $\omega_{i_0}$ is the central frequency of the idler spectrum. From energy conservation (keeping the pump frequency constant) it follows that $\partial\omega_s/\partial\omega_i=-1$ and therefore:
\begin{align}\label{Abl}
\partial \Delta k/\partial\omega_i\stackrel{!}{=} 0 = \partial k_s/\partial\omega_s-\partial k_i/\partial\omega_i=1/v_{g,s}-1/v_{g,i}.
\end{align}

\begin{figure*} [htbp]
\centering
\includegraphics[width=18cm]{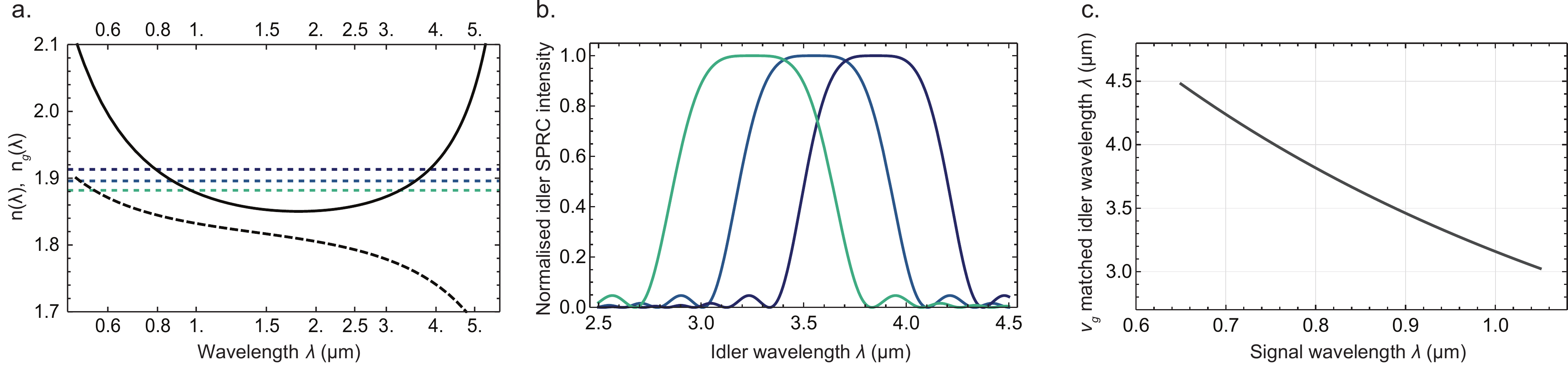}
\caption{a. Dependency of the refractive index $n$ (dashed curve) and the group index $n_{g}$ (solid curve) for the z-polarized beam in KTP. Three solutions for which $n_{g}$ is equal in the mid-infrared and in the visible or near-infrared range are highlighted with dashed horizontal lines. b. The three idler spectra corresponding to the three dashed lines in figure a for a 2\,mm long KTP crystal at room temperature without detuning of $\Delta k$ from 0 at the central idler frequency. c. Dependency of the group-velocity matched idler wavelengths on the signal wavelength for KTP.}\label{fig:KTPnGroup}
\end{figure*}
\begin{figure*} [htbp]
\centering
\includegraphics[width=12cm]{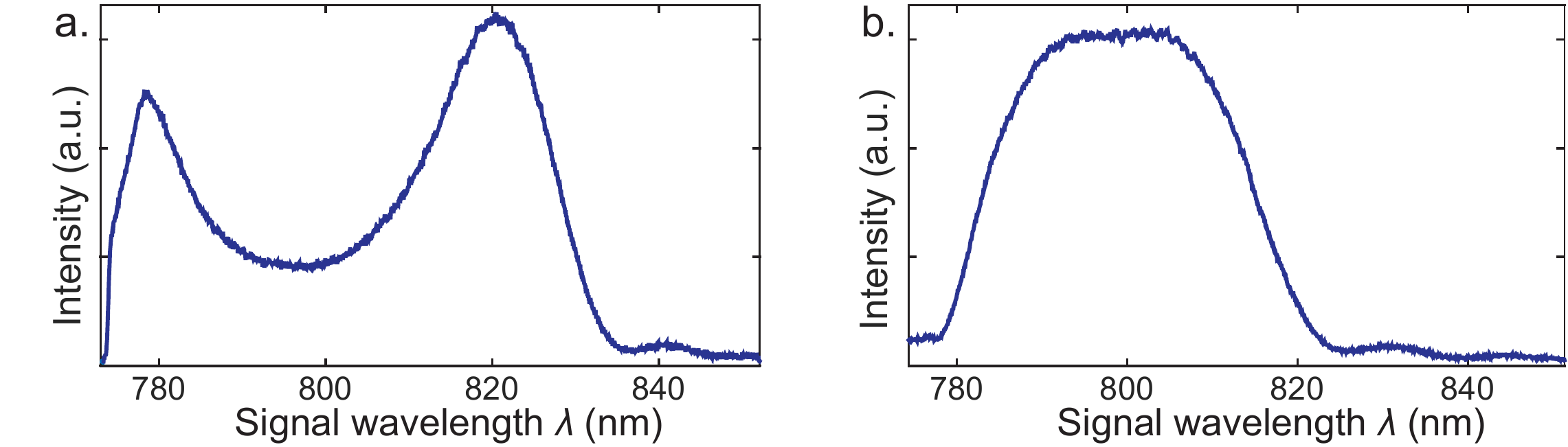}
\caption{Measured spectra with KTP poled at $\Lambda=20.00$\,\textmu m with $T=150$ $^{\circ}$C (a) and at $\Lambda=20.45$\,\textmu m with $T=40$ $^{\circ}$C (b).}\label{fig:KTPspectra}
\end{figure*}
This means that matching the group velocities $v_{g,s}$ and $v_{g,i}$ and thus the group indices of signal and idler, a large and in first order diverging bandwidth can be expected in SPDC. We call this condition signal-idler group-velocity matching.

Three equations, eq. (\ref{QPM}) for an arbitrary but fixed $\Delta k$ equal or close to 0, eq. (\ref{Abl}) and energy conservation have to be fulfilled at the central idler frequency. They contain four free parameters: $\lambda_p, \lambda_{s_0}, \lambda_{i_0}$ and $\Lambda$. This indicates that one of these can be chosen freely, allowing to find solutions for broadband nondegenerate SPDC. To demonstrate that such solutions indeed generically exist in crystals offering QPM, the cartoon in figure \ref{fig:MgOLNnGroup} depicts the typical behaviour of the refractive index and the group index of crystals such as KTP, MgO:LN and SLT. The striking feature is that the group index typically has a minimum in the infrared, allowing for matching it on both sides around it, which enables highly non-degenerate group-velocity matching solutions.

The bandwidth can be increased further by choosing the poling period $\Lambda$ such that at $\lambda_{i_0}$, which should be about in the centre of the idler spectrum, $\Delta k$ deviates slightly from 0. Due to the group-velocity matching, $\Delta k$ has an extremum or a saddle point at this wavelength. The linear term of the wavelength dependency of $\Delta k$ vanishes at this wavelength. If the second derivative does not vanish, $\Delta k$ has an extremum. Therefore, $\Delta k$ can usually be approximated by a parabola in the vicinity of this wavelength.
If $\Delta k$ has a minimum at $\lambda_{i_0}$, $\Delta k$ has to be shifted slightly below 0, if it has a maximum, it has to be moved slightly above 0. It follows that $\Delta k$ becomes 0 at one wavelength below $\lambda_{i_0}$ and at one point above $\lambda_{i_0}$ if the parabolic approximation is good enough. If the deviation of $\Delta k$ from 0 between these two points is small enough, the created power between and slightly beyond these points is nearly as high as for perfect phase matching, which increases the bandwidth further.
\begin{table*}[htbp]
\centering
\caption{Broadband phase matching results in \textmu m for SLT, MgO:LN and KTP with crystal length $L=2$\,mm}
\begin{center}
\begin{tabular}{rrrrrrrrrrr}
\hline
\multicolumn{1}{r}{crystal} &
\multicolumn{1}{l}{$\lambda_{i_0}$} & \multicolumn{1}{l}{$\lambda_{i_{\rm min}}$} & \multicolumn{1}{l}{$\lambda_{i_{\rm max}}$} & \multicolumn{1}{l}{$\Delta \lambda_{i}^{^{\rm FWHM}}$} & \multicolumn{1}{l}{$\lambda_{s_0}$} & \multicolumn{1}{l}{$\lambda_{s_{\rm min}}$} & \multicolumn{1}{l}{$\lambda_{s_{\rm max}}$} &
\multicolumn{1}{l}{$\Delta \lambda_{s}^{^{\rm FWHM}}$} &\multicolumn{1}{l}{$\lambda_{p}$} & \multicolumn{1}{l}{$\Lambda_{\rm poling }$} \\ \hline
SLT & 3.3 & 2.87 & 3.75 & 0.872 & 0.995 & 0.961 & 1.042 & 0.081 & 0.765 & 22.04 \\
& 3.8 & 3.38 & 4.24 & 0.858 & 0.866 & 0.846 & 0.891 & 0.045 & 0.705 & 19.80 \\
& 4.25 & 3.83 & 4.69 & 0.860 & 0.776 & 0.763 & 0.792 & 0.029 & 0.656 & 17.99 \\ \hline
MgO:LN & 3.3 & 2.87 & 3.74 & 0.867 & 1.094 & 1.053 & 1.151 & 0.098 & 0.822 & 23.12 \\
& 3.8 & 3.40 & 4.22 & 0.818 & 0.939 & 0.916 & 0.968 & 0.052 & 0.753 & 20.52 \\
& 4.25 & 3.86 & 4.65 & 0.791 & 0.829 & 0.816 & 0.846 & 0.030 & 0.694 & 18.34 \\ \hline
KTP & 3.25 & 2.87 & 3.64 & 0.769 & 0.986 & 0.955 & 1.028 & 0.072 & 0.757 & 24.90 \\
& 3.55 & 3.19 & 3.91 & 0.732 & 0.882 & 0.862 & 0.908 & 0.046 & 0.707 & 22.63 \\
& 3.85 & 3.51 & 4.20 & 0.696 & 0.795 & 0.782 & 0.812 & 0.030 & 0.659 & 20.45 \\ \hline
\end{tabular}
\end{center}
\label{tab:KTPtable}
\end{table*}
To show the possibility to achieve very broad bandwidths in the mid-IR by matching the group index on both sides of its minimum and using QPM, we have calculated numerically the expected down-conversion spectra for KTP, SLT and MgO:LN at room temperature.

For KTP, the results are shown in figure \ref{fig:KTPnGroup}. Three possible $v_g$-matching points are depicted in figure \ref{fig:KTPnGroup}a, the corresponding created idler spectra in figure \ref{fig:KTPnGroup}b and the $v_g$-matched signal and idler wavelengths in figure \ref{fig:KTPnGroup}c.

All calculations are for type-0 phase matching with the highest nonlinearity, i.e. all three interacting beams have z polarization (extraordinary for MgO:LN and SLT). For calculating the group-velocity matched phase-matching solutions and corresponding spectra in KTP, the Sellmeier equations by Fan et al. \cite{fan87} were used for pump and signal and those by Katz et al. \cite{Katz01} for the idler. For MgO:LN, \cite{Gayer} was used, for SLT \cite{Bruner03}. Due to the lack of Sellmeier equations for long wavelengths, some of the calculations for MgO:LN and SLT exceed the guaranteed validity range of the Sellmeier equations by at most 0.65\,\textmu m in the infrared. We assume that they still give reasonable results in these regimes. The Sellmeier equation for MgO:LN is valid up to 4.0\,\textmu m, the formula for SLT up to 4.1\,\textmu m. All calculations were done for room temperature and a crystal length of 2\,mm and without detuning of $\Delta k$ from 0 at the central idler wavelength.

For KTP and SLT, the minimum of the group index is at 1.81\,\textmu m, for MgO:LN it is at 1.92\,\textmu m. The signal has to be below this wavelength, the idler above. Examples for broadband phase matching solutions can be found in table \ref{tab:KTPtable}. Here we detail the signal and idler bandwidths, the minimum and maximum signal and idler wavelengths, and the corresponding pump wavelengths and poling periods. For all three crystals, poling periods between 18\,\textmu m and 25\,\textmu m are typical, which is technically not problematic. Depending on the parameters, the numerical results show signal bandwidths between 29\,nm and 98\,nm and idler bandwidths between 700 and 870\,nm. Signal wavelengths between 760\,nm and 1150\,nm and idler wavelengths between 2.9\,\textmu m and 4.7\,\textmu m are typical. Typical pump wavelengths are between 650\,nm and 820\,nm. The possible idler wavelengths are limited by the absorption of the crystals which starts at 4.5\,\textmu m for KTP, 5\,\textmu m for MgO:LN and 5.5\,\textmu m for SLT \cite{Nikogosyan05}. Absorption is not included in our calculations.

To demonstrate experimentally the possibility to generate very broadband SPDC with idler light in the mid-infrared using QPM and group-velocity matching, we have tested a KTP crystal with appropriately tailored poling periods for a 660\,nm cw pump laser, see figure \ref{fig:KTPspectra}. The crystal has two adjacent poling periods, 20.00 and 20.45\,\textmu m. One can choose the poling period by shifting the crystal perpendicularly to the beam. The 20.45\,\textmu m grating generates a bandwidth of about 32\,nm in the signal between 783\,nm and 815\,nm at 40\,$^{\circ}$C, corresponding to about 730\,nm bandwidth in the idler between 3.47\,\textmu m and 4.20\,\textmu m. This agrees well with the theoretical prediction, see the bottom row in table \ref{tab:KTPtable}, and even exceeds the predicted bandwidth slightly. In case of the 20.00\,\textmu m grating, $\Delta k$ is detuned from 0 at the central idler wavelength, thus creating an even wider spectrum. It achieves a bandwidth of approximately 55\,nm in the signal between 775\,nm and 828\,nm when heated to 150\,$^{\circ}$C using an oven, thus achieving a bandwidth of about 1200\,nm in the idler between 3.25\,\textmu m and 4.45\,\textmu m, corresponding to 25\,THz.

We have presented a phase-matching technique that generates very broadband non-degenerate SPDC using signal-idler group-velocity matching. Specifically it allows pair generation with the idler in the technologically relevant mid-infrared, while the signal below 1100\,nm can be easily detected with silicon detectors. This technique can be applied to all crystals for which quasi-phase matching is standard, including KTP, LN and SLT. For a 2 mm long crystal, a bandwidth of 700\,nm to 870\,nm between 2.9\,\textmu m and 4.7\,\textmu m is typical, which we show both numerically for three crystals as well as experimentally for ppKTP.

The spectral ranges of the examples provided in table \ref{tab:KTPtable} are of particular practical relevance for measurement techniques using undetected photons \cite{Kalashnikov:2016cl}: The example idler wavelengths span part of the spectral range relevant for the identification of many plastics and gases. Mid-IR light can also penetrate strongly scattering samples like ceramics and certain plastics more deeply than VIS or NIR wavelengths, which is relevant for example for OCT \cite{israelsen18,zorin18}.

Furthermore, we anticipate that this technique can be straightforwardly extended for additional flexibility: It can also be applied for polarization combinations other than type 0 phase matching and can be tuned using temperature and angle tuning.

We expect that this phase matching scheme for SPDC will find a broad range of applications, e.g. spectroscopy and OCT with undetected photons, frequency multiplexing and quantum key distribution.

\section*{Funding Information}

This work was funded by the Deutsche Forschungsgemeinschaft (DFG) within the Emmy-Noether-Programm (RA 2842/1-1).


\bibliography{references}

\end{document}